\documentclass[journal]{IEEEtran}
\ifCLASSINFOpdf
   \usepackage[pdftex]{graphicx}
   \graphicspath{{../pdf/}{../jpeg/}}
   \DeclareGraphicsExtensions{.pdf,.jpeg,.png}
\else
   \usepackage[dvips]{graphicx}
   \graphicspath{{../eps/}}
   \DeclareGraphicsExtensions{.eps}
\fi
\begin{document}
%
\title{New version of high performance Compute Node for PANDA Streaming DAQ system}

%
%
%

\author{Jingzhou~ZHAO,~\IEEEmembership{Member,~IEEE,}
        Zhen-An~LIU,~\IEEEmembership{Member,~IEEE,}
        Wenxuan~GOND,
        Pengcheng~CAO,
        Wolfgang~Kuehn,~\IEEEmembership{Member,~IEEE,}
        Thomas~Gessler,
        Bjoern~Spruck

\thanks{Manuscript received June, 2018.}
\thanks{This work was supported in part by National Natural Science Foundation of China (11435013, 11405196), National Key Program for S$\&$T Research and Development(Grant No.2016YFA0400104)and BMBF (05P12RGFPF).}
\thanks{Jingzhou.ZHAO is with State Key Laboratory of Particle Detection and Electronics, Institute of High Energy Physics, CAS, Beijing 100049 CHINA (e-mail:zhaojz@ihep.ac.cn).}
\thanks{Zhen-An LIU is with State Key Laboratory of Particle Detection and Electronics, Institute of High Energy Physics, CAS, Beijing 100049 CHINA (e-mail:liuza@ihep.ac.cn).}
\thanks{Wenxuan GONG is with State Key Laboratory of Particle Detection and Electronics, Institute of High Energy Physics, CAS, Beijing 100049 CHINA (e-mail:gongwx@ihep.ac.cn).}
\thanks{Pengcheng CAO is with State Key Laboratory of Particle Detection and Electronics, Institute of High Energy Physics, CAS, Beijing 100049 CHINA (e-mail:caopc@ihep.ac.cn).}
\thanks{Wolfgang Kuehn is with II. Physikalisches Institut, Justus-Liebig-University Giessen, 35392 Giessen, GERMANY(Wolfgang.Kuehn@exp2.physik.uni-giessen.de)}
\thanks{Thomas.Gessler is with II. Physikalisches Institut, Justus-Liebig-University Giessen, 35392 Giessen, GERMANY(Thomas.Gessler@exp2.physik.uni-giessen.de)}
\thanks{Bjoern~Spruck is with Institute of Nuclear Physics, Johannes Gutenberg University Mainz, 55128 Mainz, GERMANY(bspruck@uni-mainz.de)}}

\maketitle

\begin{abstract}
PANDA is one of the major experiments currently under construction at FAIR/Darmstadt.  Its focus is physics with high intensity and high quality anti-proton beams with momenta up to 15 GeV/c. Event rates up to 20MHz, and a typical event size between 1.5 KB and 4.5 KB. lead to data rates as high as 200 GB/s. A trigger-less streaming DAQ system is introduced in this paper, featuring event filtering based on FPGAs and a CPU/GPU farm. The Compute Node (CN) is the central board FPGA based component in this system. A new version of the ATCA based CN  is presented. Its main features include high speed data transmission, massive data buffering capabilities to support large latency for complex decion algorithms, high performance data processing and ethernet connectivity. First test results with a prototype are presented.
\end{abstract}


%
\IEEEpeerreviewmaketitle

\section{Introduction}
%
%
%
%

\IEEEPARstart{P}{ANDA} is a next generation hadron physics detector planned to be operated at the future Facility for Antiproton and Ion Research (FAIR) at Darmstadt, Germany. It will use cooled antiproton beams with a momentum between 1.5GeV/c and 15 GeV/c interacting in the high-energy storage ring (HESR) with a hydrogen cluster jet or a high frequency frozen hydrogen pellet target reaching a peak luminosity of up to 2x10$^{32}$cm$^{-2}$s$^{-1}$\cite{ref1}, corresponding to a data rate of up to 200 GB/s. However, most of the interesting physics channels have cross sections which are several orders of magnitude smaller than the total inelastic cross section. Thus, event filtering is an integral part of the DAQ system. 

\section{New concpet DAQ for PANDA}

The PANDA experiment adopts a streaming data acquisition concept in order to allow as much flexibility as possible which the complex and diverse physics objectives of the experiment require, and also to fully exploit the high interaction rate of up to 2x10$^7$ events/s. Each sub-detector system runs autonomously in a self-triggering mode, yet synchronized with a high-precision time distribution system, SODANET. Zero-suppressed and physically relevant signals are concentrated and transmitted to a high-bandwidth computing network, featuring two layers. Layer 1 consists of FPGA based Compute Nodes (CN) designed in the ATCA standard. The modular and scalable layer features high-speed interconnects via optical links and ATCA full-mesh backplanes. 

Furthermore, large local DDR4 memories attached to each FPGA permit the implementation of algorithms with large latency. The CN layer has two functions. As a first stage, burst building and event building is performed. The data stream from each PANDA - subsystem for a single burst of antiprotons is collected and stored in a single local DDR4 memory.  Using a high-precision timing system (SODANET), data originating from individual projectile-target interactions are identified and assembled into events. In a second stage, course event filtering is performed. Here, high-level algorithms for feature extraction perform tasks such as tracking, EMC cluster recognition and particle identification. The raw data is converted into a set of 4-vectors and a first stage of event filtering is performed. The main task of this stage is the rejection of events which are not of interest for a particular experiment. 

The raw data of the remaining events is transferred to the FAIR Computing server farm, where a more refined reconstruction of the events will be performed in software running on CPUs and GPUs. After further rejection of unwanted events, the remaining data is transferred to mass storage for further offline analysis.
The data acquisition system aims for an online data reduction of factor 100 -- 1000, where the performance and the partitioning of the rejection depends strongly on the properties of the PANDA detector and the physics channels of interest.

Schematic views of the system are shown in figures 1 and 2.

\begin{figure}[!t]
\centering
\includegraphics[width=3.5 in]{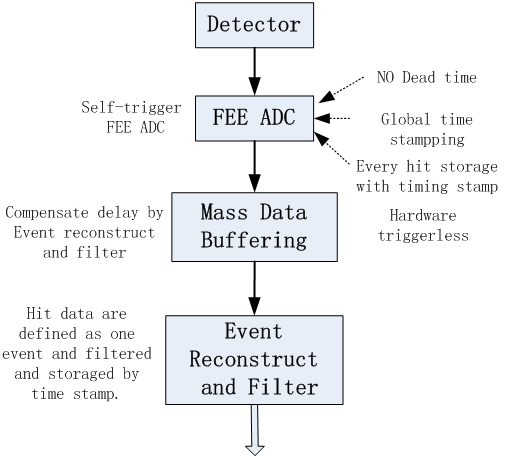}
\caption{New concept of DAQ system.}
\label{fig_sim}
\end{figure}

\section{PANDA STREAMING DAQ}
PANDA DAQ system is shown as Fig.2.\cite{ref2} It is consisted of FEE, Global timing distribution(SODANET), data concentratos and buffering, L1 network/Feature, high speed network, L2 network for final event reconstruction and filtering. 

Signals from detectors are self-triggered and sampled in FEE. The self-trigger in the FEE is based on the signal amplitude and time-stamped to allow assembly of events at a later stage. Pipelined readout mode is used to avoid dead time. 

SODANET, the global time distribution system provides absolute timing stamping for the whole system as well as data path for slow control functionality. Large data buffers are used to provide enough latency for accurate event reconstruction and event filtering.

In the FPGA based L1 network, ,information on particle momenta and particle identification are extracted. A first, more course  reconstruction of events is performed and unwanted events are rejected. A more refined event reconstruction in the L2 network using essentially offline-grade algorithms. 

For the L1 network, ATCA architecture is used, providing high speed data transmission with full-mesh backplanes, as well as scalability. The Compute Node plays a key role featuring  high performance for data transmission and processing.

\begin{figure}[!t]
\centering
\includegraphics[width=3.5 in]{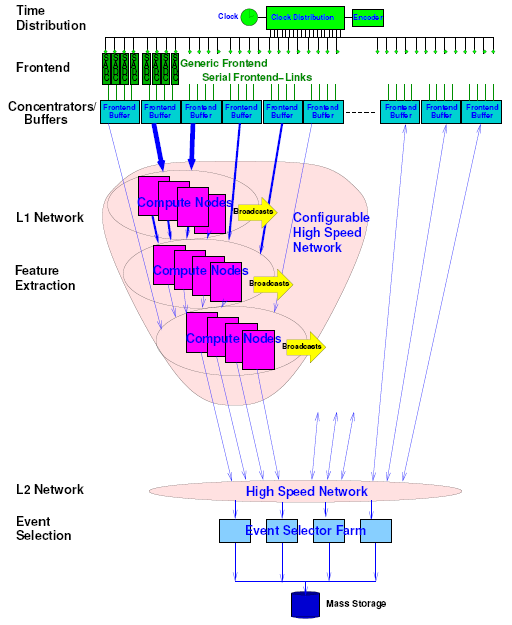}
\caption{PANDA TDAQ system structure.}
\label{fig_sim}
\end{figure}

\section{COMPUTE NODE}
Compute Node\cite{ref3} is a high speed data transmission and data processing module designed to be complaint with the ATCA standard. It consists of an ATCA carrier board CN V3, four AMC cards xFP V4 and one RTM card\cite{ref4}. CN V3 is designed based on Xilinx Virtex-4 FPGA, and features 2GB DDR2 memory, 16 MGT channels to each other slot of the backplane, 1 Gigabit Ethernet port to RTM, JTAG Hub for AMC cards, UART hub for AMCs and carrier FPGA, and ATCA compatible IPMI controller for IPMC(Intelligent Platform Management Control). xFP V4 is designed based on Xilinx Virtex-5 FX70T FPGA features 4GB DDR2, 8 MGT channels( 2 channels to SPF+ port and 6 channels to AMC connector), 1 Gigabit Ethernet port, UART port and AMC  MMC(Module Management Control).
Since the Vitex4 FPGA is is outdated, a new version of the Compute Node has been designed for the PANDA DAQ system, which is defined as CN V4.0. In the CN V4.0, the performance of CN is highly improved. The structure of CN V4.0 is shown in Fig.3. CN features a Xilinx Ultrascale FPGA xcku060 chip and up to 16 GBytes DDR4 memory. 4 single width AMC cards can be installed on the CN V4.0 carrier board. 12 backplane MGT ports are used for high speed data transmission within the shelf. The line rate of each port supports up to 16.3 Gbps. A 10 Gigabit Ethernet link is provided for data output transmission. A Gigabit Ethernet Switch connects the Ethernet ports of the  AMC cards, the ATCA Ethernet Switch, and an RTM Ethernet port switching, which can be used as slow control channel. IPMC/MMC controllers are compatible with the xTCA specification\cite{ref5}. Compute Node V4.0 is finished, as shown in Fig.4.

\begin{figure}[!t]
\centering
\includegraphics[width=3.5 in]{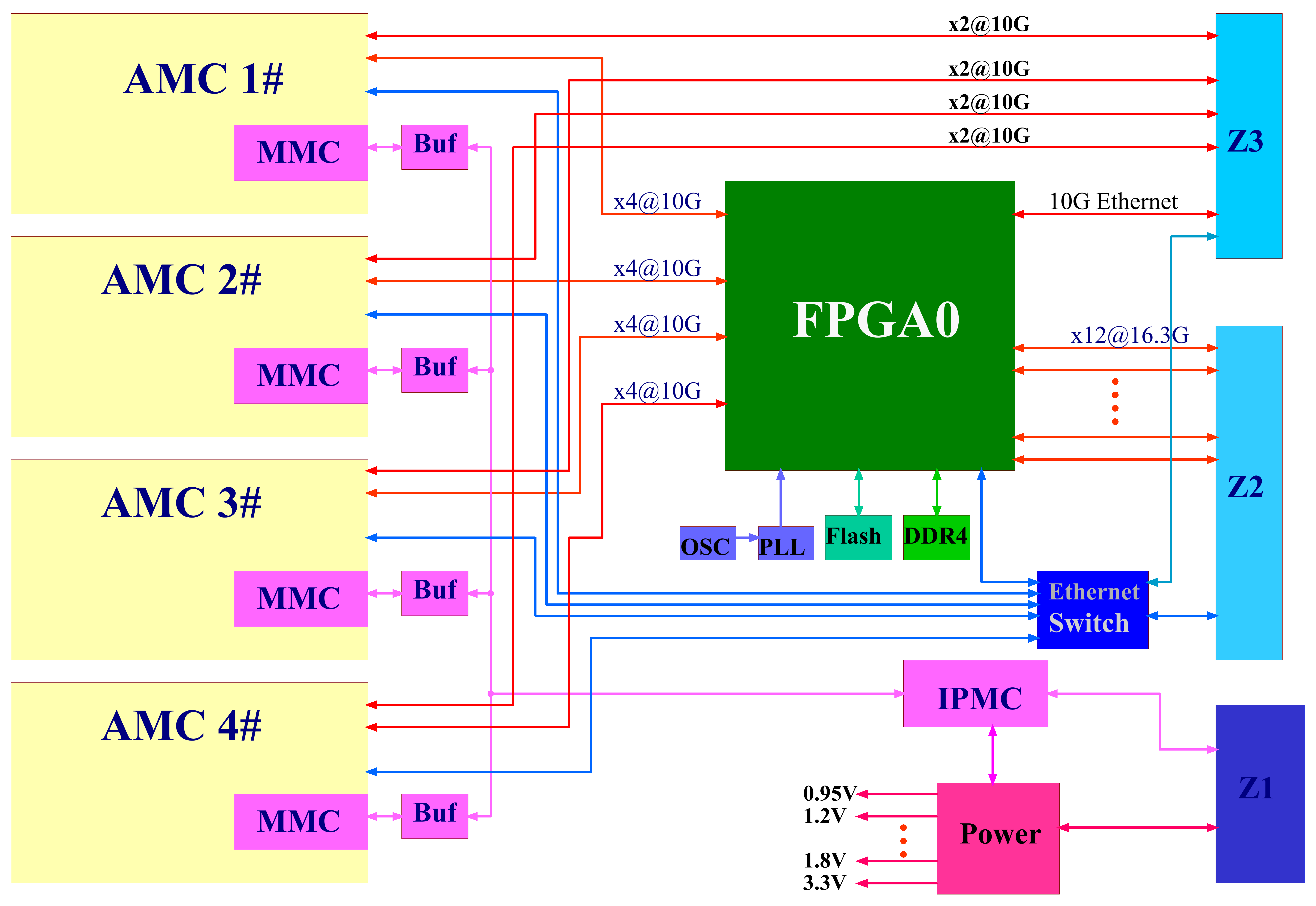}
\caption{Structure of Compute Node Version 4.0.}
\label{fig_sim}
\end{figure}

\begin{figure}[!t]
\centering
\includegraphics[width=3.5 in]{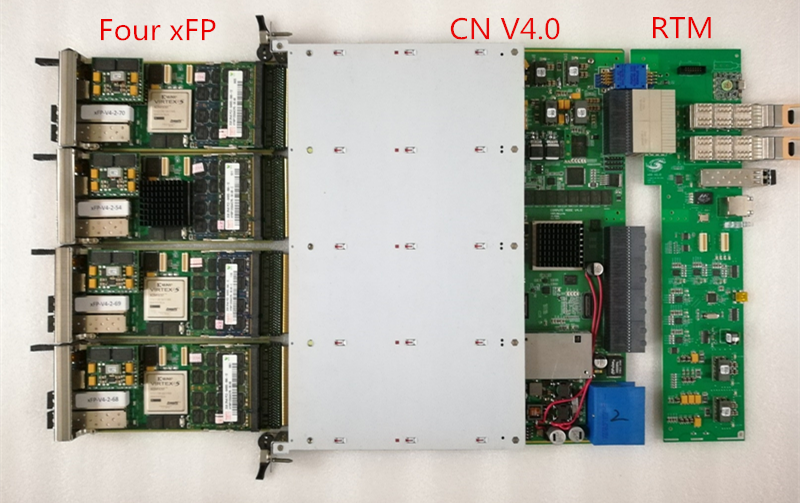}
\caption{CN V4.0 board. Carrier board and RTM card are new designed. AMC cards are xFP V4.0. CNV4.0 AMC slots are pin compatible with xFP V4.0.}
\label{fig_sim}
\end{figure}

\section{MGT TESTING RESULT}
On first prototype version of CN V4.0, speed grade 1 FPGA are used in which the MGT rate is limited to 12.5Gbps\cite{ref6}. MGT channels are tested on the  backplane channel and the AMC channels. The system comprises a Schroff AdvancedTCA 450/40 series shelf for the backplane MGT test. This shelf has two ATCA slots and its backplane and connectors support 25 Gbps per channel. 12 Backplane channels are point-to-point connected between two slots. IBERT was used for the MGT test which is the Xilinx MGT testing suite. 24 hours continuous testing was done. No error was detected on any channel. BER(Bit Error Rate) is less than 10$^{-15}$. Eye diagrams of each channel are scanned by Vivado and shown in Fig.5.

\begin{figure}[!t]
\centering
\includegraphics[width=3.5 in]{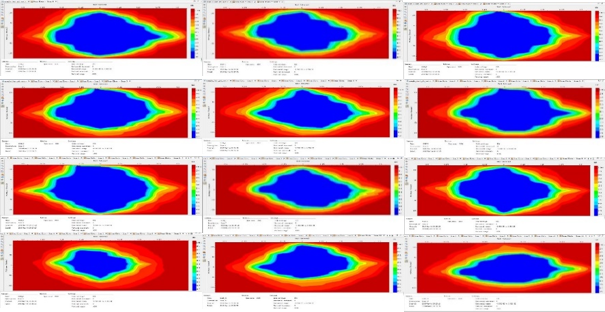}
\caption{Eye diagram for 12 Channel Backplane at 12.5Gbps.}
\label{fig_sim}
\end{figure}

CN AMC MGT channels are connections between CN carrier FPGA MGT and AMC FPGA MGT. An AMC loopback card was designed for this testing, which is as shown in Fig.6. The AMC  connector and the AMC B+ connector on CN carrier board supports only up to 10Gbps. The testing method was the same as for the CN backplane MGT test. 12 hours continuous testing was performed and BER is less than 2x10$^{-15}$. The eye diagram is shown as Fig.7.

\begin{figure}[!t]
\centering
\includegraphics[width=3.5 in]{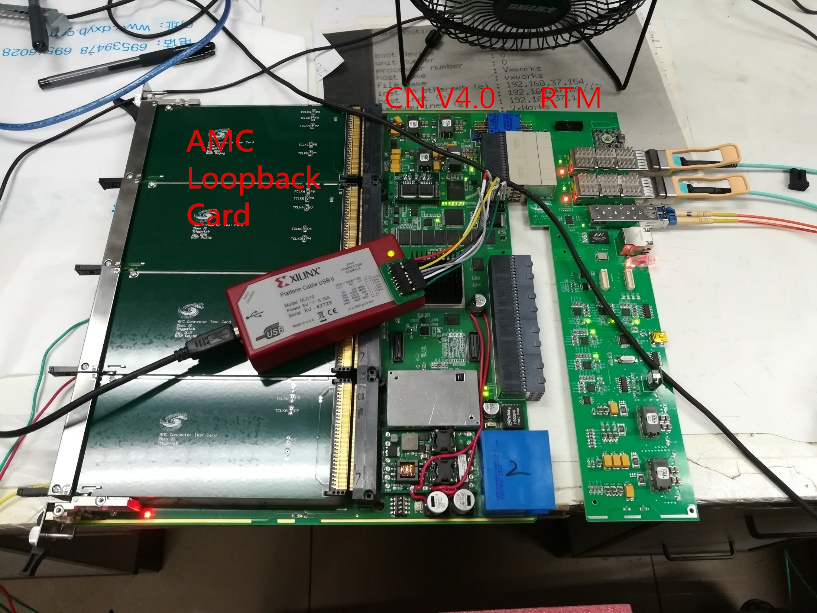}
\caption{CN AMC MGT channel testing setting up.}
\label{fig_sim}
\end{figure}

\begin{figure}[!t]
\centering
\includegraphics[width=3.5 in]{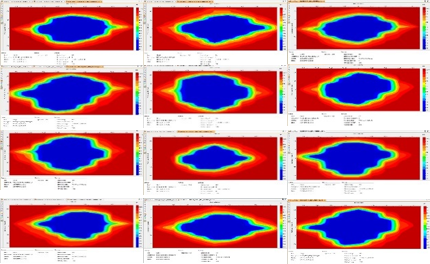}
\caption{Eye diagram for AMC MGT channel at 10Gbps.}
\label{fig_sim}
\end{figure}

\section{CONCLUSION}
The concept of a trigger-less streaming DAQ system for the PANDA experiment at FAIR/Darmstadt is presented. The Compute Node is central FPGA nased board designed to be compatible with the ATCA standard. It features  high speed data transmission, massive data buffering, high performance data processing and Ethernet connectivity. A new version of CN carrier board using Xilinx Kintex Ultrascale technology has been  designed. All essential functions are tested successfully.

\ifCLASSOPTIONcaptionsoff
  \newpage
\fi

\end{document}